\documentclass[%
reprint,
amsmath,amssymb,
aps,
]{revtex4-1}

\usepackage[normalem]{ulem}
\usepackage{graphicx}
\usepackage{dcolumn}
\usepackage{bm}
\usepackage{amssymb,amsmath,amsfonts}
\usepackage{color}

\usepackage[dvipsnames]{xcolor}
\usepackage{lineno}
\usepackage[colorlinks=true, allcolors=blue]{hyperref}

\newcommand{\mbx}{\mathbf{x}}
\newcommand{\mby}{\mathbf{y}}
\newcommand{\mbf}{\mathbf{f}}

\newcommand{\mbg}{\mathbf{g}}

\newcommand{\LL}{\mathcal{L}}
\newcommand{\LLd}{\mathcal{L}^\dagger}
\def\given{\:|\:}

\newcommand{\bi}[1]{Fig.~\ref{fig:#1}}
\newcommand{\e}[1]{\eqref{eq:#1}}
\newcommand{\mbe}{\mathbf{e}}
\newcommand{\be}{\begin{equation}}
\newcommand{\ee}{\end{equation}}
\newcommand{\lr}[1]{\left\langle #1 \right\rangle}

\begin{document}
	
	\preprint{APS/123-QED}
	
	\title{A Universal Description of Stochastic Oscillators}
	
	\author{Alberto P\'erez-Cervera}
	\email{alberp13@ucm.es}
	\affiliation{
		Universidad Complutense de Madrid. Instituto de Matemática Inderdisciplinar. Madrid, España.
	}%
	\author{Boris Gutkin}
    \email{boris.gutkin@ens.fr}
	\affiliation{Group for Neural Theory, LNC2 INSERM U960, DEC, Ecole Normale Sup\'{e}rieure - PSL University, Paris France
	}%
	\author{Peter J. Thomas}
    \email{pjthomas@case.edu}
	\affiliation{%
		Department of Mathematics, Applied Mathematics, and Statistics, Case Western Reserve University, Cleveland, Ohio.
	}%
 	\author{Benjamin Lindner}
    \email{benjamin.lindner@physik.hu-berlin.de}
	\affiliation{%
		Department of Physics \& Bernstein Center for Computational Neuroscience, Humboldt University, Berlin, Germany.
	}%

	\date{\today}
	
	\begin{abstract}
		Many systems in physics, chemistry and biology exhibit oscillations with a pronounced random component.
        Such stochastic oscillations can emerge via different mechanisms, for example linear dynamics of a stable focus with fluctuations, limit-cycle systems perturbed by noise, or excitable systems in which random inputs lead to a train of pulses.  
Despite their diverse origins, the phenomenology of random oscillations can be strikingly similar. 
Here we introduce a nonlinear transformation of stochastic oscillators to a new complex-valued function $Q^*_1(\mbx)$ that greatly simplifies and unifies the mathematical description of the oscillator's spontaneous 
activity, its response to an external time-dependent perturbation, and the correlation statistics of different oscillators that are weakly coupled. 
The function $Q^*_1(\mbx)$ is the eigenfunction of the Kolmogorov backward operator with the least negative (but non-vanishing) eigenvalue $\lambda_1=\mu_1+i\omega_1$.
The resulting power spectrum of the complex-valued function is exactly given by a Lorentz spectrum with peak frequency $\omega_1$ and half-width $\mu_1$; its susceptibility with respect to a weak external forcing is given by a simple one-pole filter, centered around $\omega_1$; and the cross-spectrum between two coupled oscillators can be easily expressed by a combination of the spontaneous power spectra of the uncoupled systems and their susceptibilities. 
Our approach makes qualitatively different stochastic oscillators comparable, provides simple characteristics for the coherence of the random oscillation, and  gives a framework for the description of weakly coupled oscillators.
		
	\end{abstract}
	
	\maketitle
	
	
\section*{Introduction}	
 
 In the age of big data, the human mind craves simple explanations of complex phenomena. The general category of “stochastic oscillations” embraces a bewildering array of natural and engineered systems in which one or more measurable quantities vary repeatedly but irregularly. 
Examples range from the molecular scale
(oscillations in genetic regulatory circuits \cite{PotWol14}) 
to the macroscopic scale
(fluctuations in predator-prey systems \cite{LugMcK08,RubEarGreeParAbb22}), 
from physical and chemical systems
(lasers \cite{DubKra99,GarCas99}, chemical oscillations \cite{GudDec96}, swaying of bridges \cite{McK14}, oscillations in aircraft wings \cite{Gib90,Lee01})
to living systems 
(oscillations in hair cell bundles \cite{BozHud03,MarHud21}, in glycolytic yeast activity \cite{HesBoi69,GolLef72}, in locomotor CPG activity \cite{Coh87}, and in cortical networks \cite{YusMac05,WalBen11}),
and from millisecond time scales 
(neuronal firing \cite{DesVin99,DzhSta04}) 
to hours 
(circadian rhythms \cite{MihHsi04,GlaSta05}) 
and longer (menstrual cycle \cite{McC71}).

A universal framework for understanding and comparing stochastic oscillations would seem to be an impossible goal, not only because nonlinear stochastic dynamical systems are intrinsically difficult to analyze, but because stochastic oscillations arise from a wide variety of underlying dynamical mechanisms.  
In the simplest case, one may obtain irregular oscillations by incorporating noise into a deterministic limit-cycle system.
Examples of noisy oscillations generated by such mechanisms include spontaneously active hair bundles in the auditory system \cite{MarBoz03}, or tonically active nerve cells in the sensory periphery, that produce trains of action potentials perturbed by ``channel noise'' (random gating of ion channels) \cite{WhiRub00,PuTho21}, or oscillations in genetic regulatory circuits perturbed by copy-number noise \cite{PotWol14}.  
In addition, there are multiple types of noise-induced oscillations: systems in which the oscillatory activity would die out in the absence of noise.  
A well-known class of noise-induced mechanisms arises when a deterministic excitable systems is perturbed by noise. Below its activation threshold, such an excitable system will not produce sustained activity.
But when perturbed by dynamical noise, an excitable system may produce an ongoing train of pulsatile activations \cite{LinGar04}.  
A nerve cell receiving a subthreshold current provides a familiar example \cite{Erm96,GutErm98,JiaGu12,FraRam23}.
Another important class of noise induced-oscillators include quasicycle systems. 
Quasicycles arise when a system has a stable equilibrium (with complex eigenvalues), perturbed by fluctuating inputs \cite{BroBre15}.  
Many physical and biological systems show random oscillations attributed to quasicycle dynamics.
Examples include underdamped linear mass-spring system immersed in a heat bath \cite{UhlOrn30},
subthreshold oscillations in nerve cells sustained by channel noise \cite{DorWhi05},  
models of EEG oscillations and intermittent cortical network activity \cite{GreMcD16,PalGei17,CabCas22,SpySap22},
and oscillations in predator-prey systems sustained by demographic (finite-population) noise \cite{LugMcK08}.
Demographic fluctuations can also sustain oscillations in systems with rock-paper-scissors interactions by yet another mechanism: noisy
heteroclinic cycle dynamics
\cite{MayLeo75,HenNgu22,RubEar23,BarTho23}.

Despite this diversity in the origins of noisy oscillations, each of the mechanisms above can be instantiated as a Markov process, for example as a system of stochastic differential equations.
Previous investigations of such systems have relied on empirical quantities such as the power spectrum (for a single unit), the cross-correlation (for multiple units), or the linear response to small-amplitude perturbations.  
The possibility of a simpler, unifying description of Markovian oscillators remains an important open question.
Ideally, one would aim to find the stochastic analogue of the well known `phase reduction'. 
In deterministic limit-cycle systems, the phase reduction \cite{Guk75,ErmTer10} (and also the phase-amplitude reduction \cite{CasGui13,WilMoe16,ShiKur17,PerSea20}) provide low-dimensional descriptions that have yielded far reaching insights into regulation, entrainment, and synchronization of oscillating systems \cite{ErmKop84,HopIzh97,PikRos02,ParHei17}. 
Although the deterministic phase concept can also be applied to some noisy systems (e.g.~single linear and nonlinear oscillators  \cite{CalHan02}  and  coupled stochastic systems \cite{TyuGol18}), generally, the notion of phase has to be generalized in a stochastic framework in order to make it applicable to cases of pure noise-induced oscillations for which a deterministic phase does not exist \cite{SchPik13,ThoLin14,BreMac18a,BreMac18b,BreMac20,PerLin21}. 
Here we go beyond such a simple extension of the phase definition, and suggest a transformation to a complex-valued function that brings about a tremendous simplification in the description of stochastic oscillators. 
We show that by transforming the system's output to a complex eigenfunction of the backward Kolmogorov operator we obtain a surprisingly simple, unified treatment of irregular oscillations,  regardless of their underlying mechanisms. 
Importantly, using our complex-valued eigenfunction description, we show that both the power spectrum and the susceptibility for single oscillators, and the cross spectrum for multiple oscillators, take dramatically simplified, universal forms.

\section*{Stochastic oscillators described by eigenfunctions}

The key step in finding a universal description comes from the observation that stochastic systems may be described not just by individual trajectories but by an \emph{ensemble} of trajectories, described by a probability density.
Nonlinear stochastic dynamical systems are difficult to analyze \cite{Str67,Ris84,HanTho82,Gar85,LinGar04}.
However, their densities evolve following  \emph{linear} dynamics, making the densities amenable to analysis as linear systems. 

We suppose that a stochastic oscillator obeys the Langevin equation
\begin{equation}
		\label{eq:SDE}
		\frac{d\mbx}{dt}=\mbf(\mbx) + \mbg(\mbx)\xi(t)
\end{equation} 
where $\xi$ represents  $k$-dimensional white Gaussian noise with uncorrelated components $\langle\xi_i(t)\xi_j(t')\rangle=\delta(t-t')\delta_{i,j}$. 
For \eqref{eq:SDE} the conditional probability of the state vector $\mbx$, given initial condition $\mbx_0$, obeys the forward Kolmogorov equation \cite{Gar85}:
\begin{align}
\label{eq:forward-eq}
		\frac{\partial}{\partial t}P(\mbx,t\given\mbx_0,s)&=\LL[P]\\
  &=-\nabla_\mbx\!\cdot\!\left( \mbf(\mbx) P \right) + \sum_{i,j} \frac{\partial^2}{\partial x_i x_j}\left(D_{ij}(\mbx)P\right)\nonumber
\end{align}
where $D=\frac12 gg^\intercal$. 
The formal adjoint of the operator $\LL$ is Kolmogorov's \emph{backward} operator $\LLd$ (also known as the generator of the Markov process \eqref{eq:SDE}, and closely related to the Koopman operator), which satisfies the equation
  \begin{align}
  \label{eq:backward-eq}
		-\frac{\partial}{\partial s}P(\mbx,t\given\mbx_0,s)&=\LLd[P]\\
  &=\mbf(\mbx_0)\!\cdot\!\!\nabla_{\mbx_0}\!\left(P \right)+\sum_{i,j}D_{ij}(\mbx_0)\frac{\partial^2 P}{\partial x_{0,i} x_{0,j}}\nonumber
\end{align}
We will assume that the operators $\LL$, $\LLd$ possess a discrete set of bi-orthogonal eigenfunctions
	\begin{equation}
	\begin{aligned}
		\label{eq:spectral-decomposition}
		\LL[P_\lambda]=\lambda P_\lambda,\quad\LLd[Q^*_{\lambda}]=\lambda Q^*_{\lambda}, \qquad \\ \langle Q_{\lambda'}\given P_{\lambda}\rangle= \int d\mbx\, Q^*_{\lambda'}(\mbx)P_\lambda(\mbx)  =  \delta_{\lambda'\lambda},
		\end{aligned}
	\end{equation}
    so that the transition probability can be expressed as \cite{Gar85} 
	\begin{equation}\label{eq:condDensity}
		P(\mbx,t|\mbx_0,s) = P_0(\mbx) + \sum_{\lambda\not=0} e^{\lambda(t-s)} P_\lambda(\mbx) Q^*_\lambda(\mbx_0),
	\end{equation}
 for $t>s$.
    That is, the transition probability $P$ can be regarded as a sum of modes, each of which decays at a rate given by the real part of its respective eigenvalue $\lambda$, leading in the long-time limit to the stationary distribution $P_0(\mbx)$.

The decaying modes in \eqref{eq:condDensity} have been shown to contain important information about the stochastic oscillation  \cite{ThoLin14, PerLin21, KatZhu21}; the most 
prominent mode being the one whose associated eigenvalue has least negative real part -- as this is the mode that decays the slowest. 
Some of us  suggested a definition of a stochastic oscillator and its stochastic phase along these lines: according to \cite{ThoLin14} the stochastic system in \eqref{eq:SDE} qualifies as \emph{robustly oscillating} if the following conditions are met: i) there exist a nontrivial eigenvalue with least negative real part $\lambda_1 = \mu_1 + i\omega_1$ which is complex valued and unique;  ii) the oscillation is pronounced, i.e.~the quality factor $|\omega_1/\mu_1|$ is much larger than one; iii) all other nontrivial eigenvalues $\lambda'$ are significantly more negative in their real parts, i.e.~$|\Re[\lambda']| \geq  2|\Re[\lambda_1]|$. 
If these conditions are fulfilled, then one can extract the stochastic asymptotic phase $\psi(\mbx)$ as the complex argument of the slowest decaying eigenfunction $Q^*_1(\mbx)$, i.e.~$\psi(\mbx) = \arg(Q^*_1(\mbx))$. 
We can then ascribe at any time $t$ a phase variable to the state $\mbx(t)$ of the system by making the nonlinear transformation to a real-valued phase of the system $\psi(t)=\psi(\mbx(t))$ (modulo $2\pi$). 
    
Here, we pursue this eigenfunction perspective further by demonstrating that the nonlinear transformation of the system, using the complex eigenfunction $Q^*_1(\mbx)$, i.e.
\be
\mbx(t) \;\; \to \;\; Q^*_1(\mbx(t)),
\ee
leads to a universal description of stochastic oscillations, independent of the specific stochastic mechanism responsible for their generation. 
The transformation to the new complex-valued variable  $Q^*_1(\mbx(t))$ entails a tremendous simplification for all of the oscillator's essential aspects. 
Firstly, we derive unifying and strikingly simple formulas for its spontaneous spectral statistics; this enables a  systematic comparison of different stochastic oscillators. 
Secondly, we also calculate its linear response to external time-dependent stimuli and derive a novel form of fluctuation-dissipation theorem. 
Thirdly, we put forward a simple but quantitatively successful theory of cross-correlations of weakly coupled stochastic oscillators. 
Hence, using the full function $Q^*_1(\mbx)$  (instead of using only its complex argument $\psi(\mbx)$) as the stochastic analog of asymptotic phase, we  achieve a true simplification and capture the universal characteristics of stochastic oscillations.

Before proceeding, we note that $Q^*_1(\mbx(t))$ has a zero stationary mean value, in the sense that
\be
\lr{Q^*_1(\mbx(t))}=\int d\mbx\, Q^*_1(\mbx) P_0(\mbx) =0,
\label{eq:mean-Q}
\ee
which follows from the bi-orthogonality relation \eqref{eq:spectral-decomposition}.
Furthermore, we normalize it to have unit variance
\be\label{eq:qNormalization}
    \lr{|Q^*_1(\mbx(t))|^2} = \int d\mbx\, |Q^*_1(\mbx)|^2 P_0(\mbx) = 1.
\ee
Finally we note that the complex argument of $Q^*_1(\mbx)$ (the above mentioned asymptotic phase of a stochastic oscillator) is only defined up to a constant phase shift. 


%


\begin{figure}[h!]
\centering
\includegraphics[width=.98\linewidth]{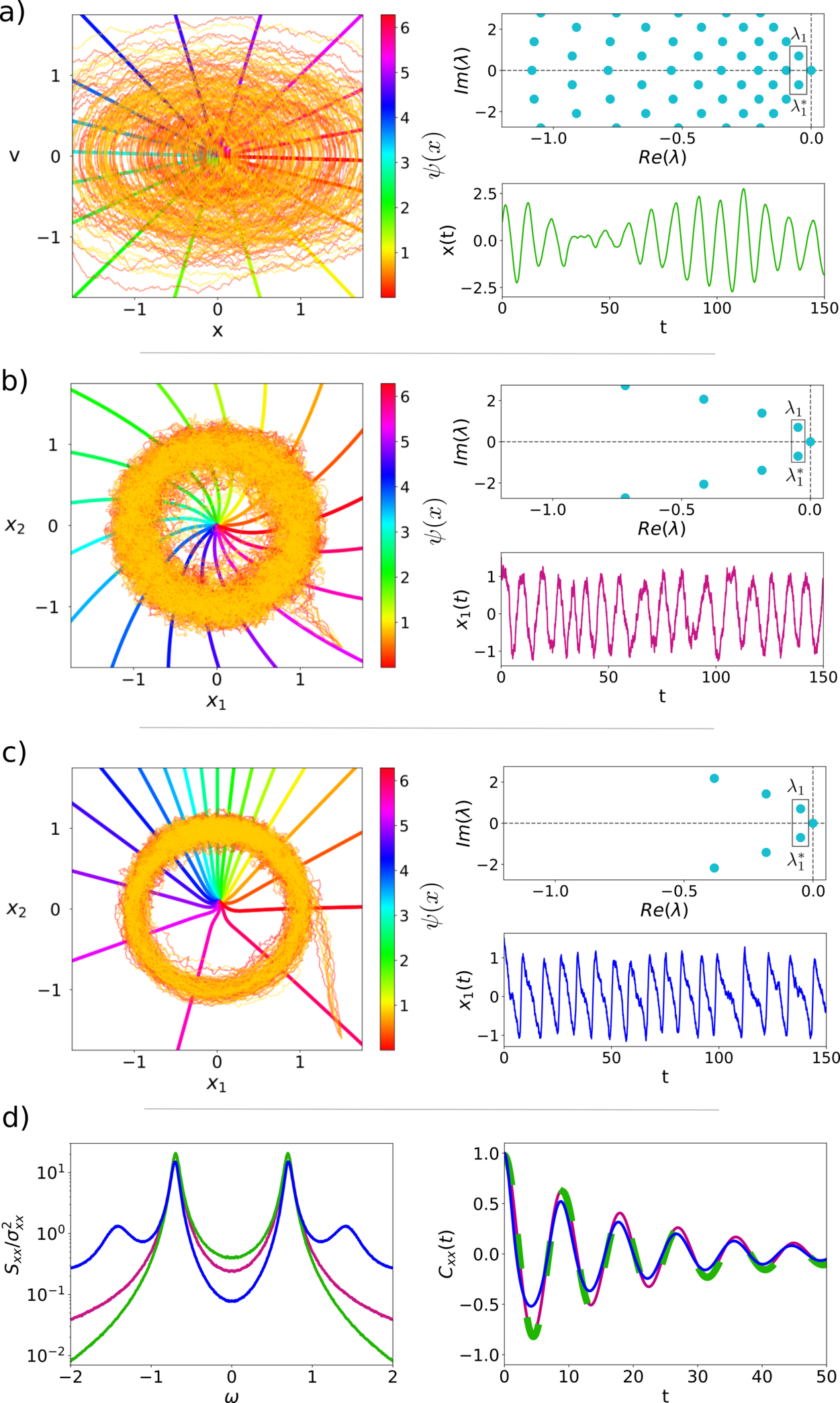}
\caption{Three models of `robust' stochastic oscillations. In the three panels  we show for each model ten sample trajectories in phase space together with the stochastic asymptotic phase $\psi(\mbx)$ (left subpanel), a time series of one of the components (lower right subpanel), and the spectrum of eigenvalues (top right subpanel). For the three models, parameters have been tuned so they have the same value for the eigenvalue $\lambda_1 = -0.048 + 0.698i$ with the smallest non-vanishing real part. \textbf{a:} Damped noisy harmonic oscillator for $\gamma= 0.096$, $\omega_0=0.699$, $D=0.01125$. \textbf{b:} Noisy Stuart-Landau for $a=1$, $b=-0.3$, $D_1=D_2=0.04$. \textbf{c:} Noisy SNIC model (beyond the bifurcation, i.e. in the limit-cycle regime) for $m=1.216$, $n=1.014$, $D_1=D_2=0.0119$.  \textbf{d:} Power spectra (left) and correlation function (right) of $x(t)$ (harmonic oscillator, green), $x_1(t)$ (noisy Stuart-Landau model, purple), and $x_1(t)$ (SNIC model, blue).}
\label{fig:coherentModels}
\end{figure}

\section*{Example models}

Throughout the paper we will illustrate our unified theory by applying it to three models
in which stochastic oscillations arise from qualitatively different mechanisms. We will use each model at two different
parameter sets -- one corresponding to a more coherent (cf.~\bi{coherentModels}) and one to a less coherent (cf.~\bi{models}) stochastic oscillation. 
We tune parameters such that all models in the more coherent case have the same leading nontrivial eigenvalue
$\lambda_1=-0.048+ i 0.698$ and thus also the same quality factor of $|\omega_1/\mu_1|=14.5$, thereby satisfying condition (ii) for a robust stochastic oscillation well. 
Likewise, we find parameters such that all models in the less coherent case have the same 
$\lambda_1=-0.168+i 0.241$ and thus also the same quality factor of $|\omega_1/\mu_1|=1.43$ which obeys condition (ii) for a robust stochastic oscillation only barely but represents the interesting limit case in which fluctuations definitely cannot be regarded as weak. 

 \emph{Damped harmonic oscillator with white noise --} As a first illustration, we consider an elementary physical model that is analytically treatable \cite{UhlOrn30}: a one-dimensional harmonic oscillator with unit mass which is subject to Stokes friction and white Gaussian noise and obeys the stochastic differential equations 
\begin{equation}\label{eq:oscDamped}
    \dot{x} = v, \qquad \dot{v} = -\gamma v - M\omega_0^2 x + \sqrt{2D} \xi(t).
\end{equation}
The model is already formulated  in non-dimensional variables (space and time) and parameters 
(friction coefficient $\gamma$, eigenfrequency $\omega_0$ and noise intensity $D$) and will be  considered exclusively in the underdamped limit ($\omega_0>\gamma/(2M)$).  We show  sample trajectories and the time courses of the stochastic oscillation for a high quality factor of $|\omega_1/\mu_1|=14.5$ in \bi{coherentModels}a and for a less coherent oscillation with $|\omega_1/\mu_1|=1.43$ in \bi{models}a. The trajectories in phase-space spend most time around the origin and in the time series of the position variable  strong  stochastic variations in amplitude and phase are seen. The eigenvalue spectra (upper right in \bi{coherentModels}a and \bi{models}a) on the left side of the complex plane are in part complex-valued but some are also purely real; the next eigenvalue to $\lambda_1$ fulfills the condition (iii) with the equal sign (the spectrum is discussed in \cite{Ris84})\footnote{Although it is well known that one cannot unambiguously define the `asymptotic phase' for a deterministic linear spiral sink, it was shown in [54] that both the $Q_1^*$ function, and hence the stochastic asymptotic phase, are well defined as long as the noise has finite amplitude.}.

\emph{Noisy Stuart-Landau oscillator --} This is the canonical model for a supercritical Hopf bifurcation, which we consider in a version endowed with white Gaussian noise
\begin{equation}\label{eq:slModel}		\begin{aligned}
	\dot{x}_1 &= a x_1 - x_2 - a(x_1^2+x_2^2)(x_1 + bx_2) +\sqrt{2D_{1}}\xi_{1}(t)\\
	\dot{x}_2 &= a x_2 + x_1 - a(x_1^2+x_2^2)(x_2 - bx_1) +\sqrt{2D_{2}}\xi_{2}(t)
\end{aligned}
\end{equation}
with $a, b \in \mathbb{R}$. 
In the absence of noise this system has a limit cycle of period $T = 2 \pi/(1 + ba)$. Because of the existing limit cycle the amplitude variations of the stochastic oscillations are much smaller than for the harmonic oscillator (see left and bottom right panels of \bi{coherentModels}b and \bi{models}b). 
The eigenvalue spectra (top right panels of \bi{coherentModels}b and \bi{models}b) are, in the displayed region, far less populated than for those of the harmonic oscillator. 
We note that there are also purely real eigenvalues outside the shown range; these are related to the amplitude of the stochastic oscillation \cite{PerLin21}.

\begin{figure}[h!]
\includegraphics[width=1\linewidth]{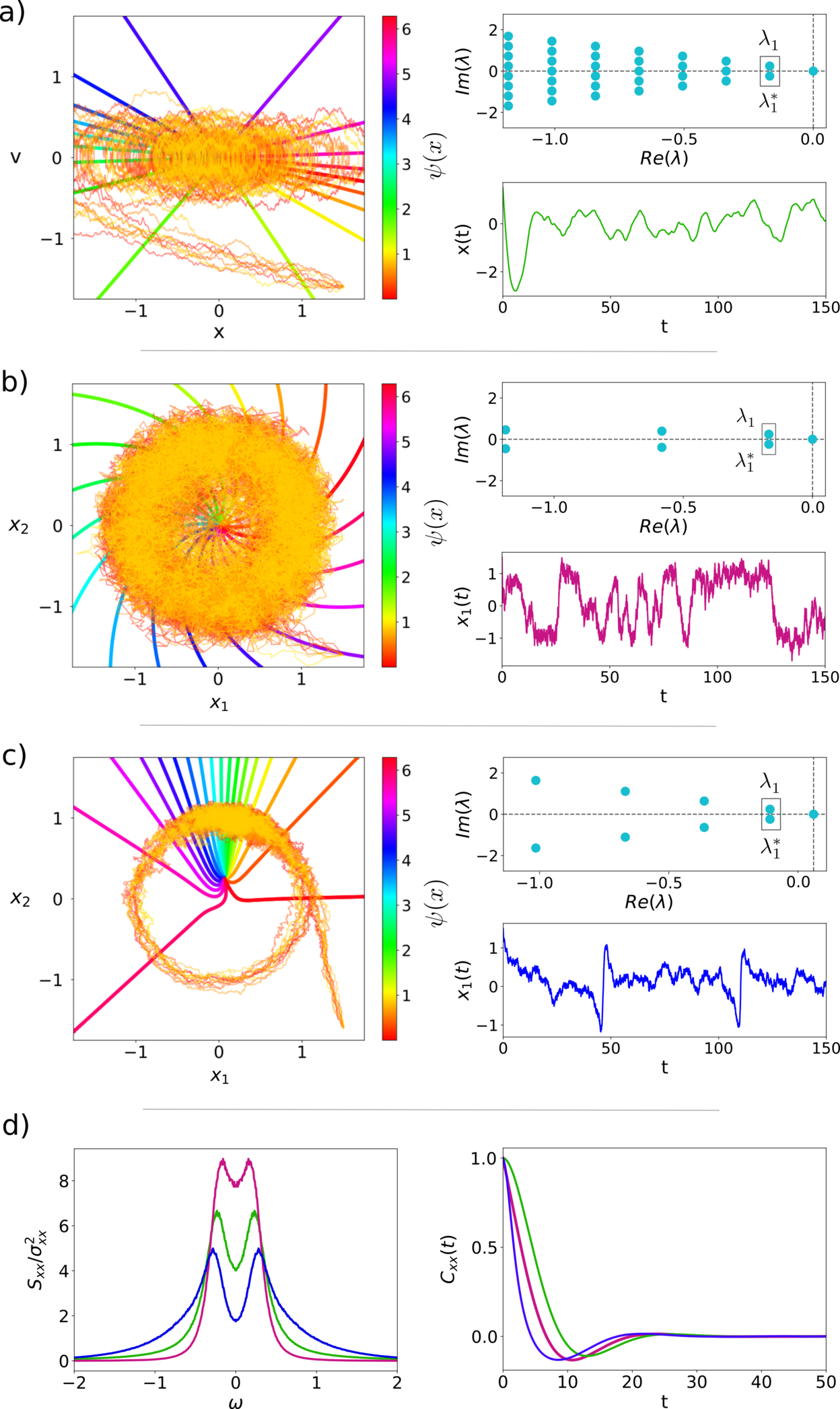}
\caption{Three models of stochastic oscillations. In the three panels we show for each model ten sample trajectories in phase space together with the stochastic asymptotic phase $\psi(\mbx)$ (left subpanel), a time series of one of the components (lower right subpanel), and the spectrum of eigenvalues (top right subpanel). For the three models, parameters have been tuned so they have the same value for slowest decaying eigenvalue $\lambda_1 = -0.168 + 0.241i$. \textbf{a:} Damped noisy harmonic oscillator for $\gamma = 0.337$, $\omega_0 = 0.294$, $D = 0.01125$. \textbf{b:} Noisy Stuart-Landau for $a = 1$, $b = -0.713$, $D_1 = D_2 = 0.0995$. \textbf{c:} Noisy SNIC model (prior to the bifurcation, i.e. in the excitable regime) for $m = 0.99$, $n=1$, $D_1 = D_2 = 0.01125$. \textbf{d:} Power spectra (left) and correlation function (right) of $x(t)$ (harmonic oscillator, green), $x_1(t)$ (noisy Stuart-Landau model, purple), and $x_1(t)$ (excitable SNIC model, blue).}
\label{fig:models}
\end{figure}

\emph{Noisy SNIC system --} A two-dimensional system that, in its deterministic version, undergoes a saddle-node bifurcation on an invariant circle (SNIC) is given by
\begin{equation}\label{eq:snicModel}
	\begin{aligned}
	\dot{x}_1 &=  nx_1 - mx_2 - x_1 (x_1^2+x_2^2) + \frac{x_2^2}{\sqrt{x_1^2+x_2^2}}   +\sqrt{2D_1}\xi_1(t),
	\\
	\dot{x}_2 &= m x_1 + n x_2 - x_2 (x_1^2+x_2^2) - \frac{x_1x_2}{\sqrt{x_1^2+x_2^2}} +\sqrt{2D_2}\xi_2(t).
	\end{aligned}
\end{equation}
Without noise, the saddle-node bifurcation from the excitable to the oscillatory regime occurs at $m=1$.
Here we consider this model endowed with white Gaussian noise once set in the oscillatory regime (leading to the more coherent stochastic oscillation, see \bi{coherentModels}c) and once set in the excitable regime (leading to the less coherent stochastic oscillation, see \bi{models}c). In marked contrast to the first two models, the $x_1$ variable of the SNIC model has a temporally asymmetric time series; however, we observe this asymmetry to be more pronounced in the excitable case.
 In this case, the trajectory stays most of the time close to the stable node and occasionally the noise causes a transition across the unstable saddle. Similarly to the Stuart-Landau case, we have fewer eigenvalues in the displayed range compared to the harmonic oscillator; again there exist purely real eigenvalues outside the range shown. 

As \bi{coherentModels}d and \bi{models}d show, despite having chosen the parameters of the three models such that they all have the same value of $\lambda_1=\mu_1+i \omega_1$ and thus share the same long-term evolution time dependence\footnote{Shared time dependence in the long-term evolution of \e{condDensity}, i.e. in  $P(\mbx,t|\mbx_0,s) \approx P_0(\mbx) +  e^{\lambda_1(t-s)} P_1(\mbx) Q^*_1(\mbx_0)$,  refers here to the shared exponential function of time; obviously, the state-dependent functions differ among the different systems.} in \e{condDensity}, the power spectra and autocorrelation functions of the models at one $\lambda_1$ differ. The differences are more pronounced for the less coherent oscillation (\bi{models}d) and they reflect the specific nature of the system. For instance, the SNIC system with its highly temporally asymmetric time series shows pronounced higher harmonics, while the harmonic oscillator does not. Except for the harmonic oscillator \cite{UhlOrn30}, it is  difficult to calculate power spectra or correlation functions for these stochastic oscillators analytically (for the Stuart-Landau oscillator, some approximations for power spectrum and linear response have been put forward in \cite{UshWue05,GleDoh06,JulDie09}).

By contrast, and as we show next, the heterogeneous profiles for the statistics of spontaneous fluctuations, as given by the power spectra or correlation functions, will be reduced to a universal form when we observe the processes through the lens of the leading backward eigenfunction $Q^*_1(\mbx(t))$.

\section*{Correlation functions and power spectra}

Generally, for any eigenfunction $Q^*_\lambda(\mbx(t))$, the correlation function is given by $C_{\lambda,\lambda}(\tau) = \langle Q^*_\lambda(\mbx(\tau)) Q_{\lambda}(\mbx(0)) \rangle$ and its Fourier transform, the power spectrum, by $S_{\lambda,\lambda}(\omega) = \int^{\infty}_{-\infty} C_{\lambda,\lambda}(\tau) e^{-i\omega \tau} d\tau$. Following \cite{Ris84} (see also SI),  we can write the autocorrelation function as an integral over the formal solution of the Fokker-Planck equation using the stationary probability density $P_0(\mbx)$ 
\begin{equation*}
    C_{\lambda,\lambda}(\tau) =  \int d\mbx\, Q^*_\lambda(\mbx)  e^{\LL(\mbx) \tau} \Big[  Q_\lambda(\mbx) P_0(\mbx) \Big].
\end{equation*}
If we expand the function $Q_\lambda(\mbx) P_0(\mbx) = \sum_{\lambda'}  \kappa_{\lambda'} P_{\lambda'}(\mbx)$ in terms of the forward eigenfunctions, and use the biorthogonal properties \eqref{eq:spectral-decomposition} of these functions (see SI), we arrive for $\tau>0$ at  
\begin{equation}\label{eq:forwardCC}
    C_{\lambda,\lambda}(\tau) = \langle|Q^*_\lambda|^2\rangle e^{\lambda \tau}.
\end{equation}
This is a strikingly simple result: the correlation function is given by the product of the stationary variance of $Q^*_\lambda$ and a complex exponential function. Specifically, for our new variable $Q^*_1(\mbx)$, taking into account \e{qNormalization} and generalizing the formula to both negative and positive time lags $\tau$, the correlation function reads
\begin{equation}\label{eq:forwardCC2}
 C_1(\tau) = 
 \exp \left[\mu_1 |\tau|  + i \omega_1 \tau \right].
\end{equation}
Real and imaginary parts of this function display damped oscillations (not shown) corresponding to the finite coherence of the stochastic oscillations. 
One characteristic of the oscillation is the quality factor $|\omega_1/\mu_1|$ that tells us how many cycles (in multiples of $2\pi$) are seen in the correlation function before the  exponential envelope  has decayed to $1/e$.

\begin{figure}[t]
\centering
\includegraphics[width=.99\linewidth]{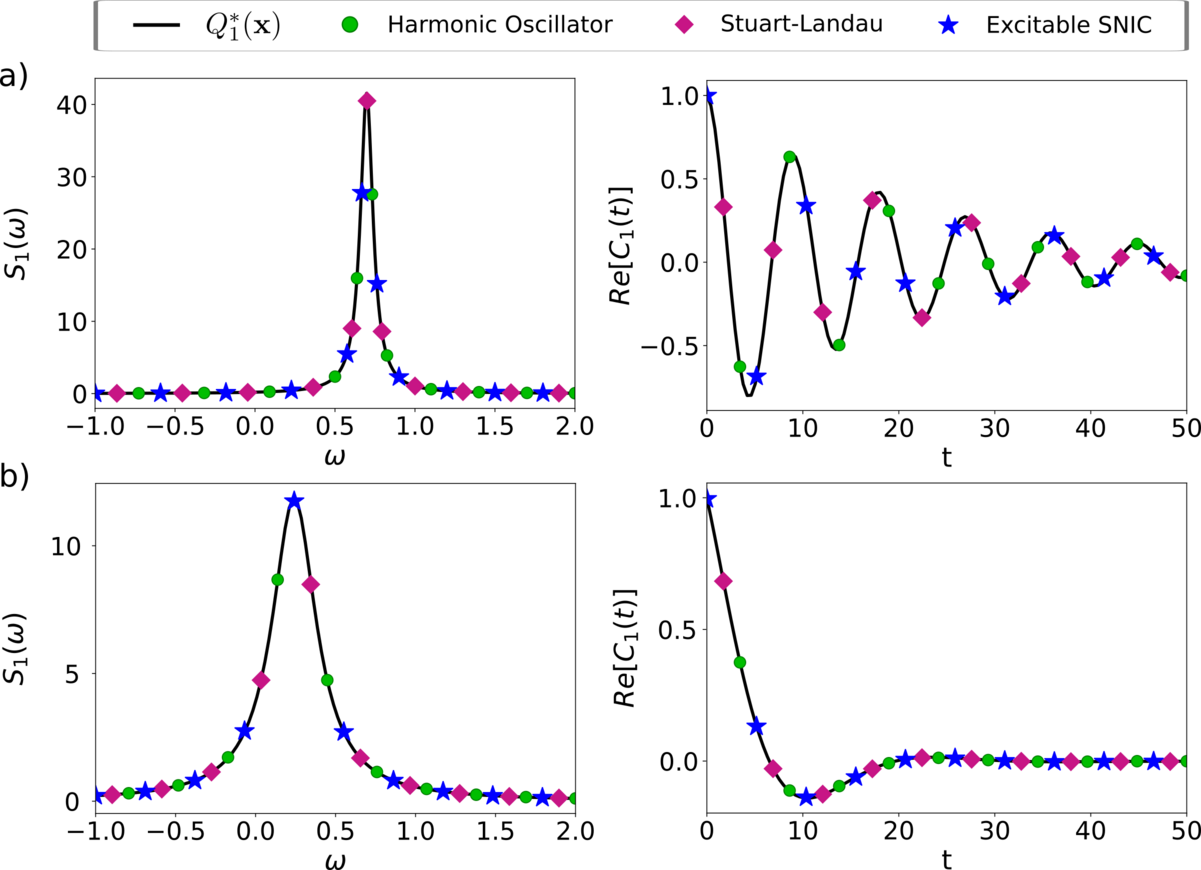}
\caption{Power spectra $S_1(\omega)$ and real part of the auto-correlation function $C_1(t)$ of $Q_1^*(\mbx(t))$ for the different models in \bi{coherentModels} (panel a) and \bi{models} (panel b). \textbf{a:} For parameters from \bi{coherentModels} (chosen such that $\lambda_1=\mu_1+i\omega_1$ is approximately the same for all models with $\mu_1=-0.048,\omega_1=0.698$, leading to a more coherent oscillation with a quality factor of $|\omega_1/\mu_1| = 14.3$) we compare \e{spectrum} (solid line), to stochastic simulations of the three models (symbols). \textbf{b:} For parameters from \bi{models} (chosen such that $\lambda_1=\mu_1+i\omega_1$ is approximately the same for all models with $\mu_1=-0.168,\omega_1=0.241$, leading to a less coherent oscillation with a quality factor of $|\omega_1/\mu_1| = 1.43$) we compare \e{spectrum} (solid line), to stochastic simulations of the three models (symbols). }
\label{fig:powerSpectra}
\end{figure}

The even simpler expression for the power spectral density corresponds to a (purely real-valued) Lorentzian, peaked at $\omega=\omega_1$ with a half-width of $\mu_1$ 
\begin{equation}
    S_1(\omega) = \frac{2 |\mu_1| }{\mu^2_1 + (\omega - \omega_1)^2}.
    \label{eq:spectrum}
\end{equation}
In \bi{powerSpectra} we show the power spectra of $Q^*_1(\mbx)$ and the real part of the auto-correlation function $C_1(\tau)$ for the parameters chosen in  \bi{coherentModels}  and \bi{models} (panels \bi{powerSpectra}a and b, respectively). As we show in Figs.~\ref{fig:coherentModels}d and \ref{fig:models}d, the power spectra and correlation functions of the models in the original variables exhibit different shapes. However, when transforming to $Q^*_1(\mbx)$, since we tuned parameters such that all three models in Figs.~\ref{fig:coherentModels} and \ref{fig:models} have the same complex eigenvalue with smallest real part, $\lambda_1$, the three very different systems possess identical power spectra. This is confirmed by our simulations (symbols) which all fall on the line predicted by \e{spectrum} (see \bi{powerSpectra}). Therefore, by means of the function $Q^*_1$ we have made the three systems quantitatively comparable and, moreover, characterizable with a simple analytical expression and two informative parameters -- the frequency and the half-width of the spectral peak (or, equivalently, the frequency and the quality factor).

By a procedure similar to that for the correlation functions and power spectra for $Q_\lambda^*(\mbx(t))$, we can also calculate expressions for the cross-correlation functions  $C_{\lambda,\lambda'}(\tau) = \langle Q^*_\lambda(\mbx(\tau)) Q_{\lambda'}(\mbx(0)) \rangle$  (see SI for details):
\begin{align}
  C_{\lambda,\lambda'}(\tau) = \lr{Q^*_\lambda Q_{\lambda'}} \begin{cases}
  e^{-\lambda'^* \tau},& \tau<0\\
   e^{\lambda \tau},& \tau>0
\end{cases}
\end{align}
and for the cross-spectra $S_{\lambda,\lambda'}(\omega) = \int^{\infty}_{-\infty} \! d\tau C_{\lambda,\lambda'}(\tau) e^{-i\omega \tau} $
\begin{equation}
\begin{aligned}\label{eq:cross-spectrum}
    S_{\lambda,\lambda'}(\omega) = - \lr{Q^*_\lambda Q_{\lambda'}} \left( \frac{1}{\lambda - i\omega} + \frac{1}{\lambda'^* + i\omega} \right).  
\end{aligned}
\end{equation}
We will need these expressions below for the theory of coupled stochastic oscillators.

\section*{Linear Response and fluctuation-dissipation theorem}
We now consider how the stochastic oscillators respond to a weak time-dependent forcing $\varepsilon p(t)$ that enters the system via a perturbation vector $\mbe$. That is, we consider 
 \begin{equation}
	\label{eq:SDE2}
	\frac{d\mbx}{dt}=\mbf(\mbx) + \varepsilon p(t) \mbe +  \mbg(\mbx)\xi(t) \qquad \mbx, \mbe \in \mathbb{R}^n.
\end{equation} 
How the time-dependent mean value of our new variable $Q^*_1(\mbx(t))$ is affected by the perturbation $p(t)$ can be described in terms of linear response theory \cite{Kub66,HanTho82,Ris84,CilGom13}
\be
\lr{Q^*_1(\mbx(t))}=\varepsilon\int_{-\infty}^t dt'\, K_{\mbe}(t-t') p(t')
\label{eq:define-LR}
\ee
where we have taken into account \e{mean-Q} and introduced the complex-valued linear-response function  $K_{\mbe}(\tau)$ (the index indicates the dependence on the direction of perturbation, $\mbe$). Equivalently, we can use the susceptibility $\chi_{\mbe}(\omega)=\int_{-\infty}^\infty d\tau e^{-i\omega \tau} K_{\mbe}(\tau)$, the Fourier transform of the response function.

In order to derive an expression for $K_{\mbe}(\tau)$, we follow Risken \cite[chapter 7]{Ris84} and express the Fokker-Planck operator by an unperturbed part $\LL$ and a perturbation part $\LL_{\mbe} = -\mbe\cdot\nabla$, leading to the Fokker-Planck equation
\begin{equation}\label{eq:pertFP}
	     \partial_t P(\mbx,t) = \big(\LL(\mbx) + \varepsilon p(t) \LL_\mbe(\mbx)\big) P(\mbx,t).
\end{equation}
Expanding the density in powers of $\varepsilon$,  $P(\mbx, t) = P_0(\mbx) + \varepsilon P_\mbe(\mbx, t) + \mathcal{O}(\varepsilon^2)$, taking only the leading linear order and expressing this by an integral over the formal time-dependent solution, we obtain
   \begin{equation}
    P_\mbe(\mbx,t) = \int^t_{-\infty} dt'\, p(t') e^{\LL(\mbx)(t-t')} \Big[\LL_\mbe(\mbx)[P_0(\mbx)]\Big]. 
\end{equation}

\begin{figure}[t]
\centering
\includegraphics[width=.98\linewidth]{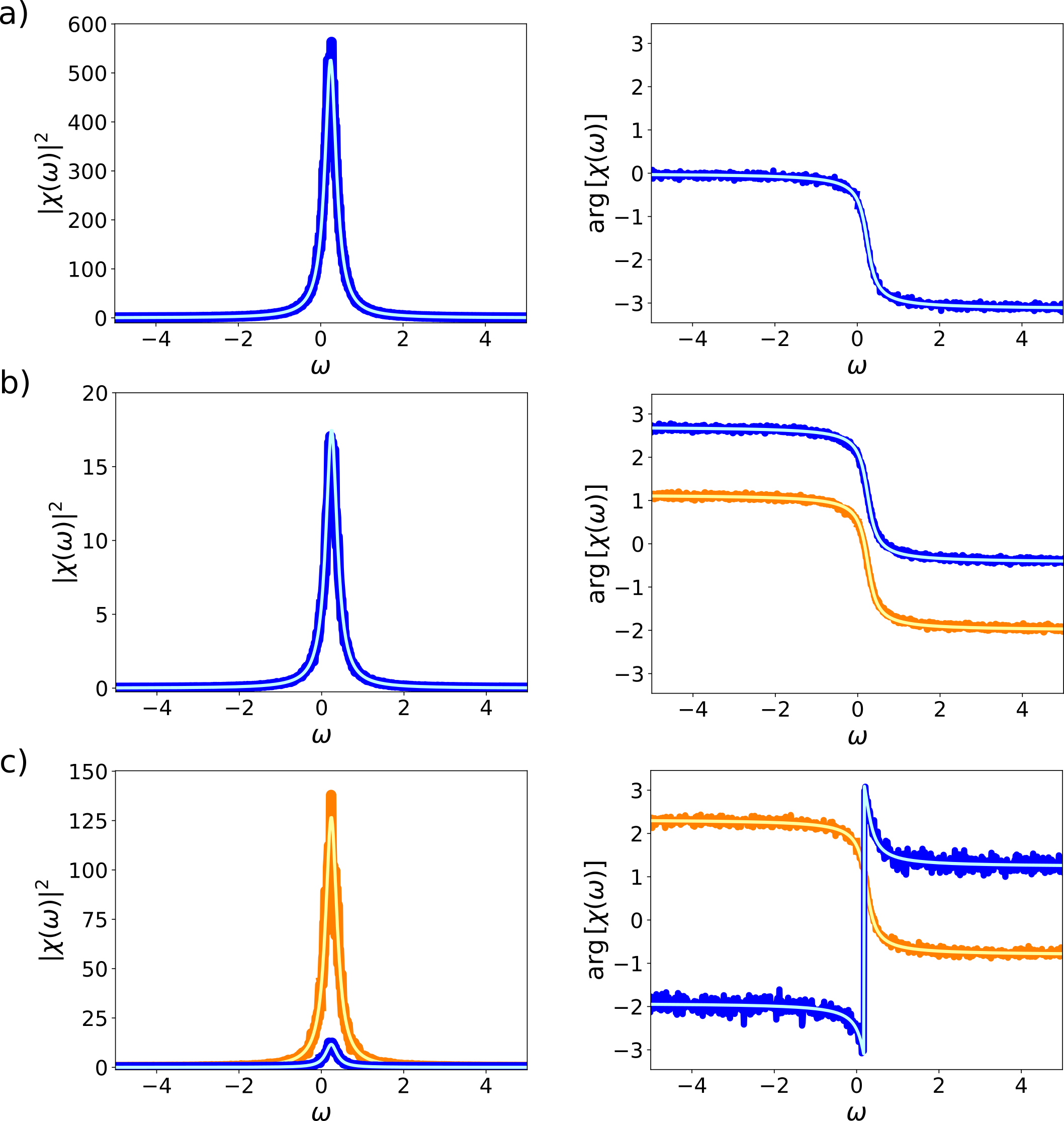}
\caption{Susceptibility functions $\chi_\mbe(\omega)$ of the variable $Q_1^*(\mbx(t))$ for the different models with the same parameters as in \bi{models} and different perturbation vectors $\mbe$ as indicated. 
For each model, we show the squared of the absolute value, $|\chi_\mbe(\omega)|^2$, (left panel) showing a Lorentzian profile and its angle $\arg(\chi_\mbe(\omega))$ (right panel). 
The perturbation vectors are $\mbe_1 = (1,0)^\top$ and $\mbe_2 = (0,1)^\top$. \textbf{a:} The harmonic oscillator $\beta_{\mbe_v} = 3.87i$ (blue, computations; cyan, theory); \textbf{b:} Stuart-Landau model $\beta_{\mbe_{1}} = -0.641 + 0.297i$ (orange, computations; yellow, theory), $\beta_{\mbe_{2}} =  -0.297 - 0.641i$, (blue, computations; cyan, theory); \textbf{c:} SNIC excitable system $\beta_{\mbe_{1}} = -1.38 - 1.3i$ (orange, computations; yellow, theory), $\beta_{\mbe_{2}} = 0.54 - 0.19i$ (blue, computations; cyan, theory).}
\label{fig:susceptibility}
\end{figure} 

By expressing the time-dependent mean value $\lr{Q^*_1(\mbx(t))}$ by the integral over $P_\mbe(\mbx,t)$ and comparing to \e{define-LR}, we obtain for the linear-response function
the intermediate result
\begin{equation}\label{eq:linearResponse}
    K_\mbe(\tau) = \int d\mbx\; Q^*_1(\mbx) e^{\LL(\mbx)\tau} \Big[\LL_\mbe(\mbx)[P_0(\mbx)]\Big], \;\; \tau>0.
\end{equation}
We expand $\LL_\mbe(\mbx)[P_0(\mbx)]=\sum_{\lambda'} \beta_{\mbe,\lambda'} P_{\lambda'}(\mbx)$ into forward eigenfunctions, use the eigenvalue equations and the  biorthogonality relation \e{spectral-decomposition}, 
and finally take into account causality (which implies $K_\mbe(\tau)\equiv 0$ for $\tau<0$) to arrive at a simple expression for the linear-response function (see SI for details)
\begin{equation}\label{eq:response-function}
    K_\mbe(\tau)  = \beta_{\mbe}\begin{cases} e^{\lambda_1 \tau}, & \tau>0\\ 0,& \text{else}\end{cases}, 
\end{equation}
where the complex-valued coefficient $\beta_\mbe = \beta_{\mbe,\lambda_1}$  (we omit the second index for ease of notation) is given by 
\be\label{eq:betaCoeff}
   \beta_{\mbe} = -\int d\mbx\, Q^*_1(\mbx) [\mbe \cdot \nabla P_0(\mbx)],
\ee
where $\nabla P_0(\mbx)$ is the gradient of the stationary density in our $n$-dimensional phase space. We note that for a stationary density $P_0(\mbx)$ obeying natural boundary conditions,  $\beta_\mbe =\mbe\cdot\lr{ \nabla Q_1^*(\mbx)}$, i.e.~the coefficient is related to the mean change of $Q_1^*(\mbx)$ in the direction of the perturbation.

The susceptibility of the stochastic oscillator is given by 
\begin{equation}\label{eq:susceptability}
    \chi_\mbe(\omega) = \int^\infty_{-\infty}d\tau\, K_\mbe(\tau) e^{-i \omega \tau}  =  \frac{\beta_\mbe}{-\mu_1 +i (\omega-\omega_1)},
\end{equation}
i.e.~a simple bandpass filter centered at $\omega=\omega_1$. 
Its modulus and its phase are given by,
\begin{equation}
\begin{aligned}
    |\chi_\mbe(\omega)| &= \frac{|\beta_\mbe|}{\sqrt{\mu_1^2 + (\omega-\omega_1)^2}}, \\ \arg(\chi_\mbe(\omega)) &= \arg(\beta_\mbe) + \arctan(\omega_1-\omega, -\mu_1).
\end{aligned}
\end{equation}
We confirm these results via numerical simulations of all three models in \bi{susceptibility} (see SI for details on measuring susceptibilities). 
For the harmonic oscillator, we show only the susceptibility for the physically relevant case of a perturbation of the velocity equation. For the Stuart-Landau model the susceptibilities for perturbations in the $x_1$ and $x_2$ directions are shown separately but coincide because of the symmetry of the model; the phase shifts are also the same up to a constant (we recall that the phase of our output variable $Q_1^*$ is only determined up to a constant phase). In contrast to the rotational symmetry of the Stuart-Landau oscillator, the excitable SNIC model differs in its response to perturbations in the $x_1$ and $x_2$ directions: perturbations in the $x_1$ direction are more efficient in kicking the system out of the stable fixed point and thus in evoking a response; consequently, $\chi_{\mbe_{x_1}} > \chi_{\mbe_{x_2}}$ for all frequencies. 

We note that we can calculate the response functions and susceptibilities of the higher eigenfunctions $Q^*_{\lambda'}(\mbx(t))$ in an analogous fashion, resulting in very similar formulas, \e{response-function} and \e{susceptability}.
The main differences are that (i) we have to use $\lambda'$ instead of $\lambda_1$, and (ii) in  the computation of the coefficient $\beta_\mbe$ in \e{betaCoeff}, we use $Q^*_{\lambda'}(\mbx)$ instead of  $Q^*_1(\mbx)$.

Turning back to the statistics of $Q^*_1(\mbx(t))$, we stress that the simple expressions for the autocorrelation function of the oscillator and its response function permit a simple connection between them, which can be regarded as a fluctuation-dissipation theorem (FDT). 
FDTs are relations between the spontaneous activity of a system and its response to external perturbations, and have been derived for thermodynamic equilibrium \cite{Kub66,LanLif71_StatPhys, Ris84} as well as for non-equilibrium settings \cite{Aga72, HanTho82, ProJoa09, WilSok17,Lin22,EngBor23}. 
For our broad model class, we obtain the simplest relation in the time domain as follows
\be\label{eq:fdt}
K_\mbe(\tau) = \beta_{\mbe} C_1(\tau), \quad \tau > 0, 
\ee
which, to the best of our knowledge, differs strongly from the generalized fluctuation-dissipation theorem that is based on the conjugated variable \cite{Aga72, HanTho82, ProJoa09}. 
Thus \e{fdt} constitutes a novel and simple fluctuation-dissipation theorem holding true for the general class of stochastic oscillators, most of which operate far from thermodynamic equilibrium. For relations between the power spectrum and the susceptibility  that are formally closer to the standard FDT of equilibrium systems \cite{LanLif71_StatPhys}, see SI.

\section*{Two weakly coupled stochastic oscillators}
 We now demonstrate that the transformation to the new variable $Q_1^*(\mbx)$ also allows for a simplified description of the statistics of weakly coupled stochastic oscillators. For simplicity, we consider only two coupled oscillators; however, the general method can be applied for larger systems of interacting units too.

We couple the two oscillators with the scalar functions $H_\mbx(\mbx,\mby) = H_{\mbx\mbx}(\mbx) + H_{\mby\mbx}(\mby)$ and $H_\mby(\mbx,\mby) = H_{\mbx\mby}(\mbx) + H_{\mby\mby}(\mby)$ along the directions $\mbe_\mbx$ and $\mbe_\mby$, respectively, and scale the coupling terms by a small parameter $\varepsilon$
\begin{equation}
	\label{eq:SDEcoupled}
	\begin{aligned}
	\dot{\mbx}&=\mbf_\mbx(\mbx) + \varepsilon \mbe_\mbx [H_{\mbx\mbx}(\mbx) + H_{\mby\mbx}(\mby)] + \mbg_\mbx(\mbx)\xi_\mbx(t), \\
	\dot{\mby}&=\mbf_\mby(\mby) + \varepsilon \mbe_\mby [H_{\mbx\mby}(\mbx) + H_{\mby\mby}(\mby)] +  \mbg_\mby(\mby)\xi_\mby(t).
	\end{aligned}
\end{equation} 
Here, the terms with mixed indices $H_{\mby\mbx}(\mby)$  ($H_{\mbx\mby}(\mbx)$) describe the effect of the $\mby$ ($\mbx$) oscillator on the $\mbx$ ($\mby$) oscillator; the diagonal terms $H_{\mbx\mbx}(\mbx)$ and $H_{\mby\mby}(\mby)$ can in principle be lumped into the drift terms  $\mbf_\mbx(\mbx)$ and $\mbf_\mby(\mby)$, respectively (which will then also change our $Q^*_1$ functions). Here we keep for clarity the diagonal terms as a perturbing input, such that the eigenfunctions $Q^*_{1_\mbx}(\mbx)$  and   $Q^*_{1_\mby}(\mby)$ are those of the uncoupled oscillators (see SI for a discussion of the alternative treatment of the problem).

We use the response functions \e{response-function} in a realisation-wise version 
\begin{equation}
 \label{eq:linear-ansatz}
	    \begin{aligned}
	    Q^*_{1_\mbx} \!\!&= Q^*_{1_\mbx,0} + \varepsilon\!\!\! \int\limits^t_{-\infty} \!\!\! dt' K_{\mbe_\mbx}(t\!-\!t') [H_{\mbx\mbx}(\mbx(t')) \!+\! H_{\mby\mbx}(\mby(t'))] \\
	    Q^*_{1_\mby} \!\!&= Q^*_{1_\mby,0} + \varepsilon\!\!\! \int\limits^t_{-\infty} \!\!\!dt' K_{\mbe_\mby}(t\!-\!t') [H_{\mbx\mby}(\mbx(t'))  \!+\! H_{\mby\mby}(\mby(t'))]
	    \end{aligned}
	\end{equation}
 and similarly for the other backward eigenmodes $Q^*_{\lambda'_\mbx}$ and $Q^*_{\lambda'_\mby}$ (see SI). In \e{linear-ansatz} the functions $Q^*_{1_\mbx,0}$ and $Q^*_{1_\mby,0}$ denote the spontaneous activity of the uncoupled oscillator, respectively. 
 A similar approximation (using the response function for the time-dependent mean value to approximate the realization-wise response of the system) has been successfully applied in the past to  stochastic network models of recurrently coupled spiking neurons \cite{LinDoi05,TroHu12}. 

\begin{figure}[ht]
\centering
\includegraphics[width=1\linewidth]{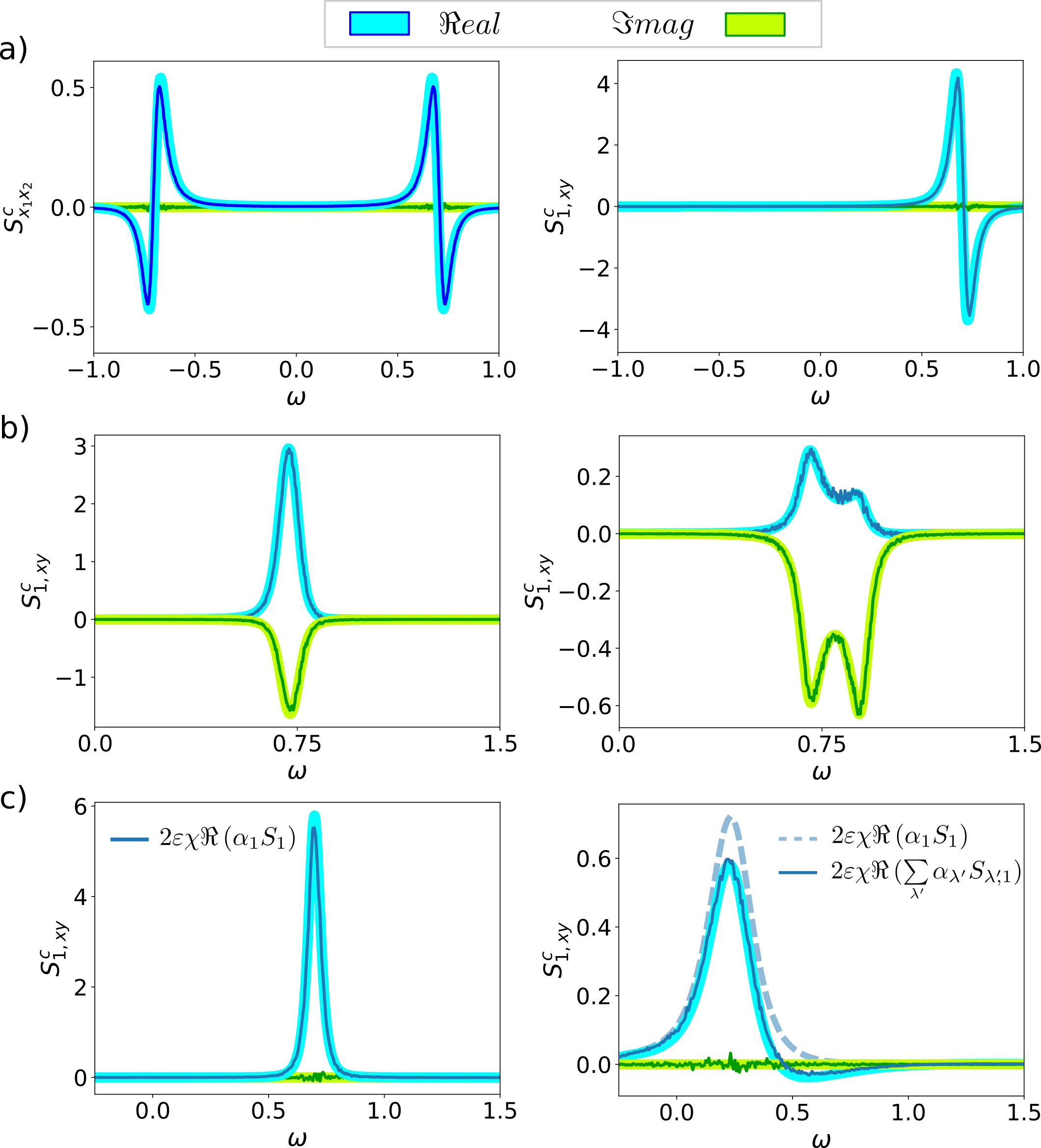}
\caption{Cross-Spectra of two coupled units for weak coupling strength $\varepsilon = 0.01$. In all panels the thin (thick) lines indicate simulations (theory); blue (green) corresponds to real (imaginary) part. \textbf{a:} For two harmonic oscillators with parameters as in \bi{coherentModels} we show cross-spectra between the position variables of each oscillator (left) and between the $Q^*_1$ functions, $S^c_{1,\mby\mbx}$. \textbf{b:} For two symmetrically coupled but non-identical Stuart-Landau oscillators we show the cross spectrum between the $Q^*_1$ functions; left: oscillators slightly detuned with one oscillator as in \bi{coherentModels} and the other one with a changed value of $b=-0.25$; right: second oscillator is more strongly detuned with $b=-0.1$. \textbf{c:} For two coupled identically SNIC systems we show the cross-spectra $S^c_{1,\mby\mbx}$ with parameters as in \bi{coherentModels} (left panel) and \bi{models} (right panel). In the right panel two versions of the theory are shown: approximations by one mode (dashed line) and by the five leading terms (solid line, see text).}
\label{fig:couplingPanel}
\end{figure} 

We assume that we can expand the coupling functions into the backward eigenfunctions as follows
	\begin{equation}
    \begin{aligned}
	   H_{\mbx\mbx}(\mbx) + H_{\mby\mbx}(\mby) &= \sum_{\lambda'_\mbx} \gamma_{\lambda'_\mbx} Q^*_{\lambda'_\mbx} + \sum_{\lambda'_\mby} \alpha_{\lambda'_\mby} Q^*_{\lambda'_\mby} \\  H_{\mbx\mby}(\mbx) + H_{\mby\mby}(\mby) &= \sum_{\lambda'_\mbx} \alpha_{\lambda'_\mbx} Q^*_{\lambda'_\mbx} + \sum_{\lambda'_\mby} \gamma_{\lambda'_\mby} Q^*_{\lambda'_\mby} 
    \end{aligned}
	\end{equation}
	where the coefficients $\gamma_{\lambda'_\mbx}, \alpha_{\lambda'_\mbx}$ are given by
	\begin{equation}\label{eq:couplingCoefs}
 \begin{aligned} 
	\gamma_{\lambda'_\mbx}\! = \!\! \int \! d\mbx\, P_{\lambda'_\mbx}(\mbx) H_{\mbx\mbx}(\mbx) , &\quad \alpha_{\lambda'_\mbx}\! = \!\! \int \! d\mbx\, P_{\lambda'_\mbx}(\mbx) H_{\mbx\mby}(\mbx) ,\\
 \gamma_{\lambda'_\mby}\! = \!\! \int \!  d\mby\,P_{\lambda'_\mby}(\mby) H_{\mby\mby}(\mby), &\quad \alpha_{\lambda'_\mby}\! = \!\! \int \! d\mby\, P_{\lambda'_\mby}(\mby) H_{\mby\mbx}(\mby).
	\end{aligned}
 \end{equation}
In addition to introducing these coefficients, we now also consider the finite-time-window Fourier transforms of the observables (see SI for details) and thus obtain from \e{linear-ansatz} 
	\begin{equation}\label{eq:6}
	    \begin{aligned}
	    \tilde{Q}^*_{1_\mbx} &= \tilde{Q}^*_{1_\mbx,0} + \varepsilon \chi_{\mbe_\mbx}  \Big( \sum_{\lambda'_\mbx} \gamma_{\lambda'_\mbx} \tilde{Q}^*_{\lambda'_\mbx} + \sum_{\lambda'_\mby} \alpha_{\lambda'_\mby}\tilde{Q}^*_{\lambda'_\mby}\Big) \\
	    \tilde{Q}^*_{1_\mby} &= \tilde{Q}^*_{1_\mby,0} + \varepsilon \chi_{\mbe_\mby}  \Big( \sum_{\lambda'_\mbx} \alpha_{\lambda'_\mbx} \tilde{Q}^*_{\lambda'_\mbx} + \sum_{\lambda'_\mby} \gamma_{\lambda'_\mby}\tilde{Q}^*_{\lambda'_\mby} \Big)
	    \end{aligned}
	\end{equation}
  and similarly for the remaining modes.
	From this linear system of equations, by a systematic expansion in the weak coupling strength $\varepsilon$ we obtain the cross-spectrum between $Q^*_{1_\mbx}$ and $Q^*_{1_\mby}$  in terms of the susceptibilities \e{susceptability} and cross-spectra between the modes of one oscillator \e{cross-spectrum} (see SI): 
	\begin{equation}\label{eq:cross-spectra-xy}
        S^c_{1,\mby\mbx} =
        \varepsilon \Big( \chi^*_{\mbe_\mbx}  \sum_{\lambda'_\mby} \alpha^*_{\lambda'_\mby} S_{1_\mby, \lambda'_\mby} +   \chi_{\mbe_\mby} \sum_{\lambda'_\mbx} \alpha_{\lambda'_\mbx} S_{\lambda'_\mbx, 1_\mbx} \Big)
\end{equation}
From this formula we can extract the following information. 
First of all, for weak coupling, the cross-spectrum between oscillators is proportional to $\varepsilon$. 
Secondly, the first term in the parenthesis consists of the susceptibility of the $\mbx$ oscillator, and a weighted sum of cross-spectra between the different eigenfunctions of the $\mby$ oscillator with the most important term being the power spectrum $S_{1_\mby}$. The complex-valued coefficients of this sum are determined by $H_{\mby\mbx}(\mby)$, the coupling function from $\mby$ to $\mbx$ (see \e{couplingCoefs}). The second term in the parenthesis is similar, only the roles of $\mbx$ and $\mby$ are switched. For two statistically identical oscillators with symmetric coupling, the second term is the complex conjugate of the first one and hence the cross-spectrum will be real-valued; any non-vanishing imaginary part thus reflects a heterogeneity in the oscillators or the coupling. For the interesting case of a purely unidirectional coupling from $\mby$ to $\mbx$, for instance, the second term in the parenthesis in \e{cross-spectra-xy} will simply vanish.

Our result for the cross-spectrum of the oscillators \e{cross-spectra-xy}  still contains an infinite sum of terms. However, as all of our numerical examples below show, just a few terms in the sums will effectively contribute. Specifically, for the case of coherent stochastic oscillators with similar frequencies,  we may restrict the sums just to the first terms (involving the spectra associated with $\lambda_1$ and $\lambda_1^*$) and still obtain accurate results.

We start testing our formula for the cross-spectrum of two identical harmonic oscillators that are weakly coupled by a spring (\bi{couplingPanel}a). In this case, of course, the cross-spectrum between the original position variables of the two oscillators can be  easily calculated and is shown in the left panel (and calculated in the SI): a purely real function with a positive lobe for $\omega<\omega_0$, a negative lobe for $\omega>\omega_0$ and everything mirrored at negative frequencies. When computing the cross-spectrum of the new variables $Q^*_{1_\mbx}$, $Q^*_{1_\mby}$, we can take advantage of the analytical expression for $Q^*_1$ (see SI), to find that the coupling function is exactly given by a linear combination of $Q^*_1$ and $Q_1$ (hence, higher coefficients $\alpha^*_{\lambda'_\mby}$ and $\alpha_{\lambda'_\mbx}$ in the expansion are identically zero; see SI for further details). Therefore, the infinite sum in \eqref{eq:cross-spectra-xy} reduces to
\begin{equation}
\label{eq:cross-spectrum-reduced}
        S^c_{1,\mby\mbx} \approx
        2\varepsilon \Re\Big( \chi^*_{\mbe} \alpha^*_{1} S_{1}   + \chi_{\mbe}^* \alpha_{1} S_{1,1} \Big),
\end{equation}
where we have omitted the $\mbx$, $\mby$ dependences of the functions on the r.h.s. since both oscillators are assumed to be identical.

As we observe in \bi{couplingPanel}a right panel, \eqref{eq:cross-spectrum-reduced} displays an excellent agreement with numerical simulations. 
If we compare the cross-spectrum of the $Q^*_{1_\mbx}$, $Q^*_{1_\mby}$ functions
with the cross-spectrum in the original position variables, we note that they look very similar with the only difference being that in $S^c_{1,\mby\mbx}(\omega)$ everything happens exclusively at positive frequencies (we consider rotating pointers in the complex plane instead of real-valued time series) and the zero crossing of the function is at $\omega_1$ (here close to $\omega_0$). Hence, the cross-spectrum of the two systems described in terms of the backward eigenfunctions reflects  the \emph{inter}dependence of the two systems appropriately. We note that, while the largest contribution to $S^c_{1,\mby\mbx}$ in \eqref{eq:cross-spectrum-reduced}  is given by the power spectra term $S_1$, the additional term $S_{1,1}$, even if it is small, has to be included to match the asymmetry between the two lobes (sizes of minimum and maximum are slightly different); including only power spectra terms would result in a strictly odd function with respect to $\omega=\omega_1$.

Next, we employ our formula \eqref{eq:cross-spectra-xy}, to study a case of symmetrically coupled but \emph{non-}identical oscillators. We consider two different noisy Stuart-Landau oscillators  diffusively coupled by their first coordinates $x_1$ and $y_1$. We  study two cases to inspect how inhomogeneities of the oscillators affect the cross-spectrum:  we set parameters such that (i) both oscillators are slightly detuned  ($\lambda_{1_\mbx}=-0.048 + 0.698i$, $\lambda_{1_\mby}= -0.047 + 0.748i$) and (ii) oscillators are more strongly detuned ($\lambda_{1_\mbx}=-0.048 + 0.698i$, $\lambda_{1_\mby}= -0.047 + 0.9i$). 
As all quality factors in this example are small and the system is rotationally symmetric (which according to our numerical observations implies $S_{1,1}(\omega)\equiv 0$), we expect that  the cross-spectrum is approximately given by
\begin{equation}\label{eq:cross-spectra-further-reduced}
        S^c_{1,\mby\mbx} =
        \varepsilon \Big( \chi^*_{\mbe_\mbx}   \alpha^*_{1_\mby} S_{1_\mby} +   \chi_{\mbe_\mby} \alpha_{1_\mbx} S_{1_\mbx} \Big).
\end{equation}
This formula agrees well with numerical simulations for both cases (see \bi{couplingPanel}b). We note that, as both oscillators are non-identical, the cross-spectrum has both non-vanishing real and imaginary parts. The effect of inhomogeneities is clearly seen by comparing left and right panels in \bi{couplingPanel}b. For small detuning (left panel), we observe a similar profile for the real and imaginary parts of $S^c_{1,\mby\mbx}$: a one lobe function, which is only different from zero around a narrow frequency band in the neighbourhood of both eigenfrequencies. As the detuning is small in case (i), the real part of $S^c_{1,\mby\mbx}$ is larger than the imaginary part. By contrast, in case (ii) with stronger detuning, the situation is reversed and the imaginary part has a larger absolute value than the real part; also now the two frequencies of the oscillator become visible by two distinct peaks in both real and imaginary parts. Indeed, the larger degree of inhomogeneity is not only captured by the increase of power in the imaginary part of $S^c_{1,\mby\mbx}$, but also in the appearance of two secondary peaks around the individual eigenfrequencies of each unit. 

\begin{figure}[t]
\centering
\includegraphics[width=1\linewidth]{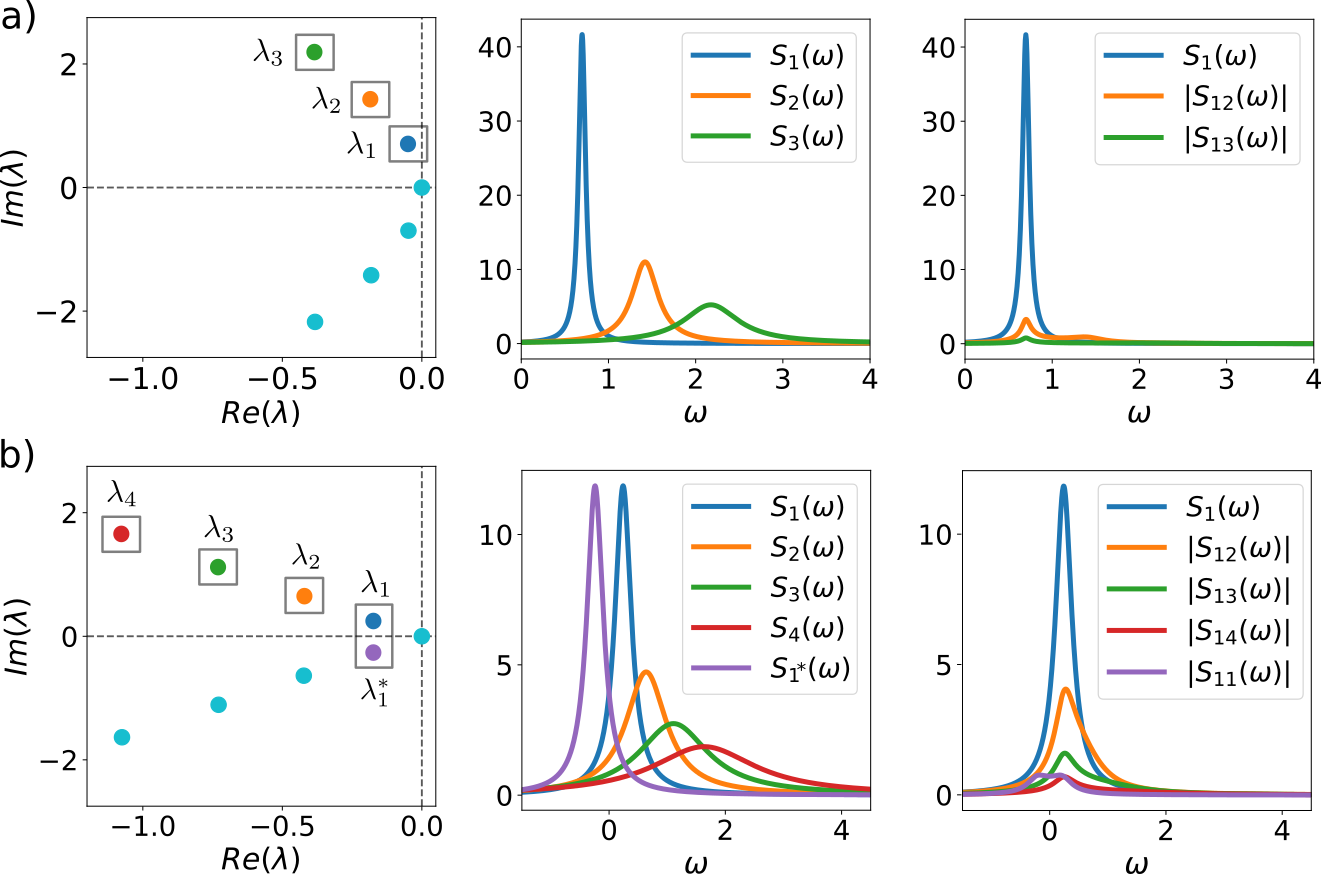}
\caption{Spectral overlap and cross-spectra 
between $Q^*_1$ and the rest of the backward modes for a more (\textbf{a}) and less (\textbf{b}) coherent oscillator. Spectrum of eigenvalues (left panels), power spectrum $S_\lambda$ of different eigenmodes (mid panels) and the cross-spectra between $Q^*_1$ and different eigenmodes (right panels). SNIC model with parameters as in \bi{coherentModels} (\textbf{a}) and as in \bi{models} (\textbf{b}).} 
\label{fig:eigenPanel}
\end{figure}

Finally, we illustrate how for less coherent oscillators, more terms in the sum are required to yield a quantitatively correct result in \e{cross-spectra-xy}. To this end, we consider first two identical SNIC systems with parameters of the more coherent case (chosen as in \bi{coherentModels}) and coupled symmetrically through their first coordinates. Here we expect
that, again, few modes are needed and, indeed, similarly to the Stuart-Landau case, we just need the power spectra term
\begin{equation}\label{eq:cross-spectra-coh-snic}
    S^c_{1,\mby\mbx} =
        2\varepsilon \Re \Big( \chi_{\mbe} \alpha_{1} S_{1} \Big)
\end{equation}
(due to symmetry, we can drop the index again and obtain a purely real-valued cross-spectrum). 
This formula shows an excellent agreement with  numerical simulations  (\bi{couplingPanel}c, left panel). However, changing the parameters to the less coherent case 
(parameters as in \bi{models}) (so now both coupled units are in the excitable regime), we find that \e{cross-spectra-coh-snic} does not suffice (compare dashed and solid curves in \bi{couplingPanel}c right panel). 
Besides the power spectra contribution, we find that obtaining an accurate prediction in this non-coherent case, requires  including cross-spectral contributions between $Q^*_1$ and the neighbouring backwards eigenmodes associated to the eigenvalues $\lambda^*_1, \lambda_2, \lambda_3$ and $\lambda_4$ (the contributions associated to $\lambda^*_2, \lambda^*_3$ and $\lambda^*_4$ have negligible covariance values).

Why did this last example require more modes than any other of the cases considered? 
There is no general answer to this question as \eqref{eq:cross-spectra-xy} depends on the coefficients $\alpha_{\lambda'_\mbx}$, $\alpha_{\lambda'_\mby}$ which depend in turn on the specific systems and the specific coupling functions. 
However, the dependence of \eqref{eq:cross-spectra-xy} on the cross-spectra $S_{1,\lambda}$, can shed some light on this question. As \eqref{eq:cross-spectrum} shows, our formula for the cross-spectra $S_{1,\lambda}$ between $Q^*_1$  and any other backward mode is weighted by their covariance $\lr{Q^*_1 Q_\lambda}$. 
The more robustly oscillatory a system is, the smaller we would expect the co-variance between $Q^*_1$ and the rest of the backward modes. The reason for such expectation is illustrated in \bi{eigenPanel}. In \bi{eigenPanel}a, we consider the SNIC model in the coherent regime and show its eigenvalue spectrum (left panel). We observe that the closest eigenvalue to $\lambda_1 = -0.048 + 0.698i$ is $\lambda_2 = \mu_2 + i\omega_2 = -0.18 + 1.42i$. Consequently, the power spectrum of $Q^*_2$, which is also given by a Lorentzian centered at $\omega_2$ and half-width of $\mu_2$, shows very little overlap with the power spectrum of $Q^*_1$ (mid panel). Hence, it is not surprising that the cross-spectrum $S_{1,2}$ is small (right panel). By contrast, if we now consider the SNIC in the less coherent case (\bi{eigenPanel}b), this scenario changes. As we see in \bi{eigenPanel}b, right panel, the eigenvalues are much closer in their imaginary parts (also real parts are larger) than in the coherent case
(now $\lambda_1 = -0.168 + i0.241$,  $\lambda_2 = -0.42 + i0.64$ and $\lambda_3 = -0.73 + i1.11$). Therefore, there is an effective overlap between their respective power spectra (mid panel) leading to non-negligible cross-spectra between $Q^*_1$ and its neighbouring modes (right panel) and these contributions have to be taken into account in the theory.  



\section*{Summary and discussion}
In this paper we have developed a simplifying framework for stochastic oscillators that can be described by systems of stochastic differential equations. 
By mapping the system's $n$-dimensional state vector to a complex-valued oscillator 
given by the eigenfunction $Q^*_1(\mbx)$ of the backward Kolmogorov operator to the eigenvalue $\lambda_1=\mu_1+i \omega_1$ with the least negative real part, we achieve a significant reduction in complexity. 
By using the transformed variable  $Q^*_1(\mbx)$, i.e.~the pair $(\Re[ Q^*_1(\mbx(t))],\Im[Q^*_1(\mbx(t))])$, 
we accomplish three major simplifications.
First, we can describe the single oscillator's spontaneous activity by a simple correlation function consisting of a single  exponential, or, equivalently, by a Lorentzian power spectrum with frequency $\omega_1$, half-width $\mu_1$, and quality factor $|\omega_1/\mu_1|$. Second, we can quantify the response to an external stimulus with a simple linear response function of the form $K(\tau)\propto \Theta(\tau)\exp(\lambda_1 \tau)$, a function that is related to the correlation function by a simple proportionality.
This result constitutes a fluctuation-dissipation theorem for a non-equilibrium system that is distinct from other theorems that have been derived in the past (e.g.~\cite{Aga72,ProJoa09,Lin22}). 
Third, by mapping the oscillator state to the $Q^*_1(\mbx)$ function, we can predict the form of the cross-correlations of coupled noisy oscillators.

We illustrated the working of the general theory by three models  that have distinct mechanisms for generating  stochastic oscillations; mathematically speaking, these were a linear system with a stable focus driven by fluctuations, the canonical model for a supercritical Hopf bifurcation endowed with noise, and a system with a saddle-node on invariant circle bifurcation likewise with uncorrelated noise. 
It is important to note that the first and the third example would not perform any oscillations (at least in the long-time limit) in the absence of noise. 
These oscillations are noise-generated in both cases, though by different mechanisms.
The second system constitutes a limit-cycle system perturbed by noise and thus here the effect of the fluctuations are easier to grasp: more noise will reduce the phase coherence of the oscillation. 
Although the three systems are very different in their dynamical mechanism, they become similar and, moreover, comparable when viewed through the lens of the   $Q^*_1(\mbx)$ function. 
Even in our framework, we still see characteristic aspects of the system, when we look at their response to external stimuli or the cross-correlation statistics for coupled systems. 

While results on the theory of the single power spectrum and the linear response are exact, the case of coupled stochastic oscillators required a new idea for the analytical calculation;
we used an ansatz that employs linear response theory (proceeding as if the dynamics in the new variable were linear). We illustrated the resulting expressions for a number of numerical examples: for the three models, for identical and non-identical oscillators, for  rather coherent and for more noisy oscillators. In all cases and for sufficiently weak coupling we found excellent agreement between the predicted and the simulated cross-spectra of the $Q_1^*$ variables  of the two systems. We take this as an indication that the true dynamics of the new variable is effectively linear. 
The reasons why this is so merit further exploration.

The universal description of stochastic oscillations put forward here, may also be used to better highlight the characteristic differences between the different systems. 
Given that two oscillators have the same $\lambda_1$ (i.e.~the same quality factor), what sets them apart? 
Should we combine the information for the leading complex-valued eigenvalue and its eigenfunction with that of the first purely real-valued eigenvalue and the associated eigenfunction, which can be used to define the stochastic limit cycle \cite{PerLin22}?  
Or should we rather compare the higher oscillatory modes $Q^*_{\lambda'}$ (with $|\Re(\lambda')|>|\mu_1|$ and $\Im(\lambda')\neq 0$), that play such a prominent role in our theory of coupled oscillators? 
It seems to us that both comparisons offer a novel perspective for the finer categorization of stochastic oscillators.  

The relation between our universal description of stochastic oscillators and the classical phase description of deterministic oscillators bears further discussion. 
Recall that, in deterministic limit-cycle systems, the phase can be obtained from the argument of the principal eigenfunction of the Koopman operator with purely \textit{imaginary} eigenvalue \cite{MauMez18, ShiKur20}. 
Upon introducing noise into the system, this eigenvalue develops a negative real part. 
Indeed it becomes $\lambda_1$ and its associated eigenfunction becomes $Q^*_1$. 
This connection between the deterministic phase and the $Q^*_1$ function in the noise vanishing limit is not coincidental since the Kolmogorov backward operator $\LLd$ corresponds to the stochastic version of the Koopman operator \cite{CrnMac20}. 
Indeed, the relationship between the stochastic asymptotic phase (the complex argument of $Q^*_1$) and the deterministic phase in the limit $D\to0$ has  already been noted \cite{ThoLin14,EngKue21, KatZhu21,KatNak23}. 
Hence, our transformation also embraces the deterministic case and  connects cleanly with the well-established deterministic  Koopman-operator framework \cite{KamKai20}. 

Returning to the specific results of our paper, we note that they can be generalized  in different directions. 
First of all, even if our general setup includes multiplicative noise, for simplicity we restricted all of our examples to Langevin systems with additive white Gaussian noise. 
Nothing keeps us from finding the eigenfunction to the eigenvalue with the least negative real part and to make the transformation to this complex-valued variable in a system with multiplicative noise. 
Likewise, we are not restricted to systems with Gaussian white noise but can also apply the method to Markov processes described by a master equation (for which there exist also a backward operator with eigenfunctions; one such example has been already treated in \cite{ThoLin14} for the extraction of the asymptotic phase of a stochastic neuron model with discrete channel noise). 
More generally even, any jump-drift-diffusion process \cite{Gar85} described by a master equation (with additional drift and diffusion terms) that shows the hallmarks of robust stochastic oscillations, can be captured by our universal description in terms of the $Q^*_1(\mbx)$ function. 
Our formulas for the main characteristics will not change and, for instance, the power spectrum of such systems in the new variable will still be a pure Lorentzian, the response function a pure exponential, etc. 
Another straightforward generalization concerns the external perturbation: this could (and will in certain cases) also depend on the state of the system. 
This will mainly affect the definition of the complex-valued coefficient $\beta$ that appears in the response function.

An exciting challenge is to extend our analysis of two coupled oscillators to the general case of $N$ weakly coupled oscillators with its obvious applications to neural  \cite{Bru00,LycErm10,KhaFum22}, mechano-sensory \cite{FruJue14,WitMan20,RooFab21}, genetic \cite{PurSav10,PotWol14}, metabolic \cite{RowSte21}, and energy supply networks \cite{KueThr19}, to name but a few examples. 
Because our analytical approach can be generalized to this case, different scenarios of connectivity (sparse, random or structured) and heterogenity (in the single oscillator properties or in the connections) can be studied analytically. 
Moreover, the summed activity of subgroups of oscillators at the mesoscopic level can be calculated from the cross-spectral statistics of single stochastic oscillators. 

Our theory was here applied to stochastic models, but applications to data are conceivable.
In \cite{ThoLin14} it was demonstrated how the stochastic asymptotic phase (the complex argument of the function $Q^*_1$) can be extracted from data.
In the same way, the function $Q^*_1$ itself can be found, provided the data are consistent with a robustly oscillatory Markov process. 
We suggest that if one were to propose a method for extracting either $Q_1^*$, or higher modes $Q_\lambda^*$, our results offer a test: whether the resulting power spectra fit simple Lorentzians at the respective eigenfrequencies (see SI).
Thus, in light of the direct link between the Kolmogorov backwards operator and the stochastic Koopman operator, our work may help advance methods for extracting Koopman eigenfunctions from data \cite{Sch10, BudMar12, ProBru16, MauMez20}, as well as for providing physical interpretations of particular modes \cite{ThoLin14, PerLin21}.
%
%

Our framework also offers a test of a key assumption, namely that the stochastic oscillation arises from a Markov process.
Markovianity is an important characteristic of stochastic processes, and different methods to test for it are currently under debate (see e.g. \cite{BerMak18,LapGod21,EngBor23}). 
For the important class of stochastic oscillators, computing the statistics of $Q^*_1(\mbx(t))$, and specifically probing for a purely Lorentzian line shape, may provide another independent tool to test for Markovianity.  

In summary, there are many open problems that can be studied within the framework put forward here.

\section*{Supplementary Information (SI)}

The SI for this manuscript can be accessed \href{http://people.physik.hu-berlin.de/~lindner/PDF/Perez-etal-2023-SI.pdf}{here}

\begin{acknowledgments}
APC acknowledges funding by the Bernstein Center for Computational Neuroscience Berlin
for a research stay in Berlin in fall 2022, where the core ideas of this paper were developed.  PJT acknowledges support from NSF grant DMS-2052109, and from the Oberlin College Department of Mathematics.
\end{acknowledgments}

	\bibliography{apssamp}

\end{document}